\documentclass[10pt,doublecolumn,twoside]{IEEEtran}
\usepackage{cite}
\usepackage{amsmath,amssymb,amsfonts}
\usepackage{algorithmic}
\usepackage{graphicx}
\usepackage{textcomp}
\usepackage{epstopdf}
\usepackage{float}
\usepackage{esint}
\usepackage{subfigure}
\usepackage{color}
\usepackage{mathrsfs}
\usepackage{algorithm}
\usepackage{algorithmic}
\usepackage{bm}
\usepackage{cases}
\usepackage{changes}

\newtheorem{lemma}{Lemma}

\def\BibTeX{{\rm B\kern-.05em{\sc i\kern-.025em b}\kern-.08em
    T\kern-.1667em\lower.7ex\hbox{E}\kern-.125emX}}

\makeatletter

\newcommand{\Rmnum}[1]{\expandafter\@slowromancap\romannumeral #1@}
\makeatother

\newcommand{\ls}[1]
    {\dimen0=\fontdimen6\the\font
     \lineskip=#1\dimen0
     \advance\lineskip.5\fontdimen5\the\font
     \advance\lineskip-\dimen0
     \lineskiplimit=.9\lineskip
     \baselineskip=\lineskip
     \advance\baselineskip\dimen0
     \normallineskip\lineskip
     \normallineskiplimit\lineskiplimit
     \normalbaselineskip\baselineskip
     \ignorespaces
    }

\begin{document}
\title{Statistical Age-of-Information Optimization for Status Update Over Multi-State Fading Channels}

\author{Yuquan Xiao,~\IEEEmembership{Graduate Student Member,~IEEE}, and Qinghe Du,~\IEEEmembership{Member,~IEEE}
\thanks{\ls{.5}Copyright (c) 2015 IEEE. Personal use of this material is permitted. However, permission to use this material for any other purposes must be obtained from the IEEE by sending a request to pubs-permissions@ieee.org. This work was supported in part by the National Key R\&D Program of China under Grant No. 2020YFB1807700, in part by the National Natural Science Foundation of China under Grant 62071373,  in part by the Key Research and Development Program of Shaanxi Province under the Grant No. 2023-GHZD-05, and in part by the Innovation Capability Support Program of Shaanxi under the Grant No. 2021TD-08. The correspondence author of this paper is Qinghe Du (e-mail:duqinghe@mail.xjtu.edu.cn). Yuquan Xiao and Qinghe Du are with the School of Information and Communications Engineering, Xi'an Jiaotong University, Xi'an 710049, China.}
\vspace{-27pt}
}

\maketitle
\begin{abstract}
Age of information (AoI) is a powerful metric to evaluate the freshness of information, where minimization of average statistics, such as the average AoI and average peak AoI, currently prevails in guiding freshness optimization for related applications. Although minimizing the statistics does improve the received information's freshness for status update systems in the sense of average, the time-varying fading characteristics of wireless channels often cause uncertain yet frequent age violations. The recently-proposed statistical AoI metric can better characterize more features of AoI dynamics, which evaluates the achievable minimum peak AoI under the certain constraint on age violation probability. In this paper, we study the statistical AoI minimization problem for status update systems over multi-state fading channels, which can effectively upper-bound the AoI violation probability but introduce the prohibitively-high computing complexity. To resolve this issue, we tackle the problem with a two-fold approach. For a small AoI exponent, the problem is approximated via a fractional programming problem. For a large AoI exponent, the problem is converted to a convex problem. Solving the two problems respectively, we derive the near-optimal sampling interval for diverse status update systems. Insightful observations are obtained on how sampling interval shall be tuned as a decreasing function of channel state information (CSI). Surprisingly, for the extremely stringent AoI requirement, the sampling interval converges to a constant regardless of CSI's variation. Numerical results verify effectiveness as well as superiority of our proposed scheme.
\end{abstract}




\vspace{-10pt}
\section{Introduction}
Wireless status update systems, where the time-varying status of the source is continuously sampled, packaged, and delivered to the destination via wireless channels, benefit our lives everywhere, such as smart home systems, industrial automation, emerging metaverse, etc. How to guarantee the information freshness for these applications is then ever-increasingly essential. Age of information (AoI) is a powerful metric to evaluate the freshness of information~\cite{kaul_real-time_2012}, which records the time-elapsing process within the time window starting from generation of each status sample and ending upon its arrival at the destination. Correspondingly, optimizing the statistics of AoI to control and manage resource allocation in wireless status update systems attracts a great deal of research attention~\cite{Huang2022Age,Zhao2023Improved}.

Both average AoI and average peak AoI metrics have been widely studied towards resource management for wireless transmissions in the status update systems. The authors of~\cite{kadota_scheduling_2018} studied the sum of weighted average AoI minimization problems and gave the near-optimal scheduling policy in broadcast wireless networks. The authors of \cite{jiang_peak_2021} researched the average peak AoI minimization problems for unmanned-aerial-vehicle (UAV) aided wireless sensor networks, where the UAV trajectory and transmit power are both tuned to achieve the minimum peak AoI. Literature \cite{champati_minimum_2021} analyzed the achievable minimum average peak AoI for a single-server-single-source queuing system with a specified service time distribution. However, the metrics defined by the average statistics, like average AoI and average peak AoI, cannot effectively evaluate, control, and manage the tail probability of AoI. A large tail probability of AoI could result in poor users' experience in many applications, like metaverse, industrial automation, self-driving cars, etc.



Aside from average statistics of AoI, the maximum peak AoI, which is desired to be minimized as much as possible, is another metric to impact the resource optimization for transmissions. For example, the authors of~\cite{cao_2020_peak} studied the minimization of maximum peak AoI for UAV-aided relay transmissions. Although lowering the maximum peak AoI metric imposes a hard bound on AoI without exception, due to the uncertainty of wireless channel, the attained maximum peak AoI might be hard to confine within a low level in many cases, thus often failing to satisfy users' experiences.

Above unsolved issues motivate the study on shaping the distribution rather than simply control a single average or peak statistic of AoI. Some literatures investigate how to minimize the violation probability of age exceeding a certain threshold~\cite{Statistical2021Champati,Optimized2020Abdel}. Similarly, there are also several works focusing on minimizing the age threshold or its upper bound, e.g., the conditional value-at-risk of age, with the constrained violation probability~\cite{Zhou2020Risk,Hu2021Status}. For instance, the authors of~\cite{Statistical2021Champati} studied the problem of minimization of the age violation probability for wireless two-hop control systems, where a heuristic sampling rate scheme was derived by a mathematically relaxation way. The authors of~\cite{Zhou2020Risk} investigated the multi-objective optimization problem in real-time monitoring systems, where the conditional value-at-risk of age was considered and solving which the history-dependent updating policy was obtained. However, the minimization problems with the objective of either age violation probability or conditional value-at-risk of age are sometimes complex and difficult to tackle. Alternatively, the recently-proposed statistical AoI metric can effectively evaluate the achievable minimum peak AoI in a statistical yet rational way, i.e., imposing a certain age violation probability constraint \cite{zhang_statistical_2021}. This metric has the fairly desired mathematical properties and the sophisticated optimization methods can be then used. Specifically, the statistical AoI is defined as the normalized log-moment generating function (LMGF) of peak AoI with a required AoI exponent, which indicates the exponentially decaying rate of the probability of peak age exceeding a certain threshold. In the meanwhile, the statistical AoI functions as an unified metric integrating the average statistic. In particular, when the AoI exponent decreases to zero, the statistical AoI reduces to the average peak AoI \cite{costa_age_2016}. When the AoI exponent goes up to infinity, the statistical AoI converges to the maximum peak AoI \cite{cao_2020_peak}.

While statistical AoI injects new vigor into wireless status update systems, how to make use of this metric to optimize dynamic sampling process and resource management to fulfill diverse applications' requirements has neither been sufficiently understood nor adequately researched. Toward this end, we use the concept of the statistical AoI to study the achievable minimum peak AoI over wireless fading channels with the given age violation probability constraint. Specifically, the main contributions are summarized as follows:
\begin{itemize}
  \item We formulate the statistical AoI minimization problem over the multi-state fading channels, where the given age violation probability constraint characterized by the specified AoI exponent, which can serve for the variantly information freshness-sensitive applications.
  \item The optimal solution is hard to track due to prohibitively-high complexity. We tackle the problem via a two-fold approach. For the small AoI exponent, the problem is approximated as a fractional programming problem. For the large AoI exponent, the problem is approximated as a convex problem. A near-optimal sampling interval scheme, which can respond to different AoI exponents, is derived by solving these two problems.
  \item For a given AoI exponent, we show how sampling interval shall be tuned as a decreasing function of channel state information (CSI). In particular, for the extremely stringent AoI requirement, the sampling interval converges to a constant regardless of CSI's variation. We also conduct the numerical simulation to verify effectiveness as well as superiority of our proposed scheme.
\end{itemize}




The remainder of this paper is organized as follows. Section II describes the system model. Section III formulates the statistical AoI minimization problem for status update systems over multi-state fading channels. Then, the sampling interval scheme is derived in Section IV. Section~V presents performance evaluation for our proposed scheme. The paper concludes with Section VI.

\section{System Model}
\subsection{System Description}
As shown in Fig. \ref{fig:aoi_wireless_fading}(a), we consider a wireless status update system over the multi-state fading channels. Time is divided into equal-length slots. We denote by $T$ and $B$ the duration of one slot and the channel bandwidth, respectively. The wireless fading channel has $K$ states, which is characterized by the received signal-to-noise-ratio (SNR) given the unit transmit power. We denote by $\gamma_k$ the SNR of the $k$th state, where $k\in\mathbb{K}$ and $\mathbb{K}=\{1, 2, ..., K\}$. Without loss of generality, we assume that $\gamma_1 < \gamma_2 < ... < \gamma_K$. The probability that the channel stays at the $k$th state is $p_k$. Since the transmitter and receiver in wireless status update system are often fixed in location or slowly moving, we assume that the channel coherence time, denoted by $T_c$, is much longer than the duration of one time slot. Also, the channel state is assumed to independently vary from one channel coherence time to the next channel coherence time. On the whole, time is divided into two distinct time periods, namely time slot and channel coherence time. Multiple status-sampling packets, which can each be transmitted over several consecutive time slots, could potentially be transmitted within a single channel coherence time. Our destination is to control \emph{how many samples should be scheduled under the different channel states,} i.e., to optimize the sampling interval under the different channel states to make the information received at the receiver as fresh as much.
\vspace{-10pt}
\subsection{AoI Processes at the Transmitter and Receiver}
\begin{figure}
  \centering
  \includegraphics[scale = 0.63]{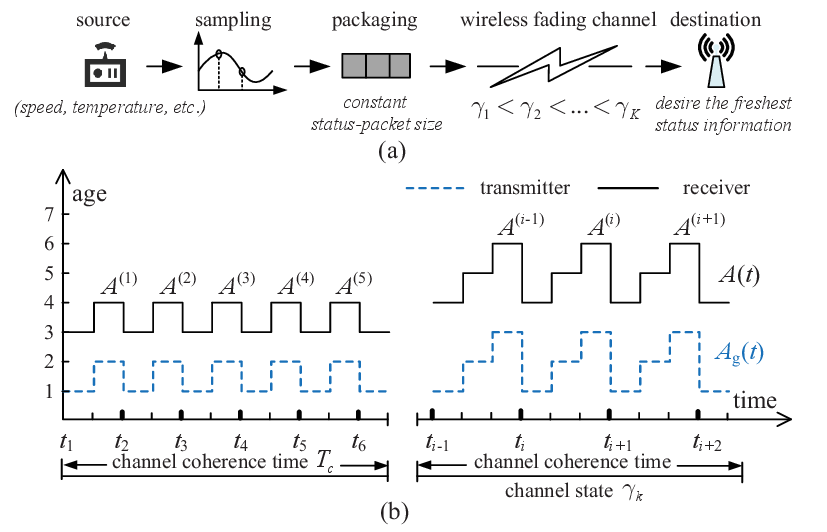}
  \caption{(a) The system model for wireless status update over the multi-state fading channels; (b) The age processes at transmitter and receiver.}
  \label{fig:aoi_wireless_fading}
\end{figure}
The generate-at-will model is used in this paper, where the real-time status packet is assumed to be generated right away once we have a willingness~\cite{Bhat2021Throughput}. After the latest status packet is just transmitted, we will immediately sample the current status and then transmit the generated status packet. Thus, the sampling interval is equal to the transmission interval. The rationale behind this arrangement is elucidated as follows. 1) By using all of time resources to transmit a constant sized status packet, we could potentially conserve a greater amount of power resources. 2) Compared with that of independent sampling interval and transmission time, this arrangement is of low complexity for the design and implementation of the system. In addition, the acknowledgement feedback and retransmission mechanism is not considered in this paper, because we do not need to deliver the previous status packet again no matter it is successfully transmitted or not. Instead, in each transmission, the freshest status packet will be delivered, which can potentially achieve the lower AoI. Suppose that during the $i$th sampling and transmission, the channel is staying at the $k$th state. We denote the number of slots used for the $i$th transmission by $n_k^{(i)}$. Then, following the standard AoI definition~\cite{kaul_real-time_2012}, the AoI process can be shown in Fig. \ref{fig:aoi_wireless_fading}(b), where $A_g(t)$ and $A(t)$ denote the AoI processes at transmitter and receiver, respectively. We do not consider the propagation delay over the air because it is much smaller than the transmission delay. As shown in Fig. \ref{fig:aoi_wireless_fading}(b), the $i$th status packet is generated from the transmitter at time instant $t_i$ and delivered to the receiver at time instant $t_{i+1}$. Once the $i$th status packet is generated, the age at the transmitter is reset to $1$. Once the $i$th status packet arrives at the receiver, the age at the receiver is updated to $t_{i+1} - t_i + 1$. At the interval of generation or arrival of the status packet, $A_g(t)$ and $A(t)$ increase linearly as time elapses. We denote by $A^{(i)}$ the $i$th peak age at the receiver. Based on the relation shown in Fig. \ref{fig:aoi_wireless_fading}(b), $A^{(i)}$ can be given by
\begin{align}
A^{(i)} = \left(n_{k}^{(i - 1)} + n_k^{(i)}\right)T,
\end{align}
where $n_{k}^{(i - 1)}$ and $n_k^{(i)}$ are the numbers of time slots used for the $(i-1)$th and $i$th transmissions, respectively. Since the transmission interval in the same channel state is identical, we have $n_{k}^{(i - 1)} = n_k^{(i)}$. Note that the notation ``$(i)$" for $n_k^{(i)}$ is omitted later in this paper for presentation convenience. Similarly, unless otherwise stated we use $A$ to represent $A^{(i)}$. Thus, the peak age can be further given by $A = 2n_k T$. Also, we require
\begin{align}\label{eq:peak_constraint}
n_k \leq n_{\rm max}
\end{align}
to avoid the extremely large peak age, where $n_{\rm max}$ represents the maximum tolerable numbers of slots for one status-packet delivery.

\subsection{Low-Power Characteristics in Status Update Systems}

The average power constraint is imposed as most status update tasks are carried out by wireless sensors with low-power transmissions. We denote by $\mu_k$ the transmit power at the $k$th channel state. Then, the average power constraint can be given as follows:
\begin{align}\label{eq:average_power_constraint}
\sum_{k = 1}^{K}\mu_k p_k \leq \bar{\mu},
\end{align}
where $\bar{\mu}$ represents the average transmit power of the wireless sensors.

\vspace{-10pt}
\subsection{Finite Blocklength Transmission}
Since the size of status packet is relatively small, the wireless status update transmissions are accordingly aligned with the short packet transmission in the finite blocklength regime. According to the finite blocklength capacity \cite{polyanskiy_channel_2010}, the achievable channel coding rate, denoted by $r(n_k)$, at the $k$th channel state, is given as follows:
\begin{align}\label{eq:channel_capacity}
r(n_k) = \log_2(1 + \mu_k\gamma_k) - \sqrt{\frac{V}{n_kTB}}f^{-1}_Q(\varepsilon)\log_2 e,
\end{align}
where $V$ represents the channel dispersion, $f^{-1}_Q(\cdot)$ is the inverse of Gaussian $Q$-function, and $\varepsilon$ is error transmission probability. Error transmission probability in the finite-blocklength regime has an impact on AoI. Large error transmission probability could often result in the failure transmission of real-time status packet and then the AoI becomes large. To weaken the influence of error transmission probability on the AoI, we in this paper demand that the value of error transmission probability in Eq.~(\ref{eq:channel_capacity}) should be set very small, i.e., less than $10^{-3}$, to guarantee reliable transmissions. We denote by $N_{\rm bit}$ the number of bits for each status packet. Thus, the number of bits which can be transmitted at the $k$th channel state with $n_k$ slots should be larger than $N_{\rm bit}$, i.e.,
\begin{align}\label{eq:bit_constraint}
r(n_k)n_k TB \geq N_{\rm bit}.
\end{align}

\section{Statistical AoI Minimization for Status Update over Multi-State Fading Channels}
In order to assess the timeliness of information in status update systems, a measure known as the average peak AoI was introduced~\cite{costa_age_2016}. This metric is effective at capturing the fundamental features of the aging process. Nevertheless, the average statistics are inadequate for capturing the extreme features of the distribution of AoI. Even if the average AoI or average peak AoI is minimized, there remains the possibility of certain incoming packets becoming outdated, leading to detrimental effects on systems involved in status monitoring, management, and control. In order to assess the uncertainty in the age process and depict the tail attributes of AoI, the employment of moment generating function (MGF) is advantageous. Furthermore, to effectively govern and control the tail probability of peak AoI, the concept of statistical AoI has been devised~\cite{zhang_statistical_2021}, which is mathematically defined as follows.

For a general AoI process $A(t)$, the statistical AoI, denoted by $\Delta_{\rm sta}(\theta)$, is defined as
\begin{align}\label{eq:effective-aoi}
\Delta_{\rm sta}(\theta) = \frac{1}{\theta}\log M_A(\theta),
\end{align}
where $M_A(\theta)$ is the MGF of peak AoI of $A(t)$, i.e., the expectation of $e^{\theta A}$, and $\theta$ is called \emph{AoI exponent}. Then, based on Chernoff-Cram\'{e}r bound, the violation probability of $A$ exceeding a threshold at the tail is upper-bounded by
\begin{align}\label{eq:Chernoff_crameir_AoI}
\mathrm{Pr}\left(A \geq \Delta_{\rm sta}(\theta) + \sigma\right) \leq \frac{M_A(\theta)}{e^{\theta\Delta_{\rm sta}(\theta) + \theta\sigma}} = e^{-\theta\sigma},
\end{align}
where $\sigma$ is an arbitrary positive value. From Eq. (\ref{eq:Chernoff_crameir_AoI}), we find that as $\theta$ changes from 0 to $\infty$, the violation probability exponentially varies from 1 to 0.
In the absence of any limitations on the probability of violation ($\theta = 0$), the statistical AoI can be equated with the average peak AoI~\cite{costa_age_2016}. Conversely, if the tightest possible restriction is enforced on the tail violation probability ($\theta = \infty$), the statistical AoI approaches the maximum peak AoI~\cite{cao_2020_peak}. Mathematically, we have
\begin{subequations}
\begin{numcases}{\hspace{-1cm}}
\lim_{\theta \rightarrow 0}\Delta_{\rm sta}(\theta) = \lim_{N \rightarrow \infty} \frac{1}{N} \sum_{i = 1}^{N} A^{(i)},\label{eq:degrade_average_peak_aoi}\\
\lim_{\theta \rightarrow \infty}\Delta_{\rm sta}(\theta) = \max\left\{A^{(i)}, i = 1, 2, ...\right\}.\label{eq:degrade_max_peak_aoi}
\end{numcases}
\end{subequations}

Based on the definition specified in Eq. (\ref{eq:effective-aoi}), the statistical AoI for $A$ is obtained as follows:
\begin{align}\label{eq:effective-aoi_fading}
\Delta_{\rm sta}(\theta) = \frac{1}{\theta}\log \sum_{k = 1}^{K}\mathrm{Pr}(A = 2n_kT)e^{2\theta n_kT},
\end{align}
where $\mathrm{Pr}(A = 2n_kT)$ is the probability distribution function (PDF) of peak age and is specified by
\begin{align}\label{eq:pdf_aoi}
\mathrm{Pr}(A = 2n_kT) = \frac{p_k}{n_k}\left(\sum_{i = 1}^{K}\frac{p_i}{n_i}\right)^{-1}.
\end{align}
The rational of (\ref{eq:pdf_aoi}) is given as follows. Based on Fig.~\ref{fig:aoi_wireless_fading}(b), because the channel coherence time is $T_c$, the number of peak age within one channel coherence time can be obtained as $\frac{T_c}{n_kT}$ at the $k$th channel state, that is, the smaller the $n_k$ is, the larger number of peak age will be within one channel coherence time. With the denotation that the probability of the $k$th channel state is $p_k$, the probability that the peak age equals to $2n_k T$ can be presented as follows:
\begin{align}
\mathrm{Pr}(A = 2n_kT) = \frac{T_c}{n_kT}p_k\left(\sum_{i = 1}^{K}\frac{T_c}{n_iT}p_i\right)^{-1}.
\end{align}
Eliminating $\frac{T_c}{T}$, we obtain Eq.~(\ref{eq:pdf_aoi}).

What we are interested in is how to optimize the sampling interval under the different channel states as well as the freshness requirement. There is a discernible balance between transmission power and information freshness. For a specified channel state, increasing the sampling interval has the potential to decrease transmit power, thus freeing up more resources for other channel states. Conversely, opting for a smaller sampling interval can yield greater information freshness. This would allow for a more efficient allocation of resources over the varying channel states and enhance overall performance, i.e., there exists the optimal sampling interval scheme for status update. To seek this scheme, the optimization problem, denoted by \textbf{\emph{P}0}, is formulated as follows:
\begin{subequations}
\begin{align}
\textbf{\emph{P}0}:
\displaystyle \min_{n_k,\mu_k} \,\,\,& \Delta_{\rm sta}(\theta),\\
{\rm{s.t.}}\quad
& (\ref{eq:peak_constraint}), (\ref{eq:average_power_constraint}), \mbox{and } (\ref{eq:bit_constraint}).
\end{align}
\end{subequations}

\section{Efficient Sampling Interval Scheme for Status Update}
\subsection{Problem Transformation}
The objective of \textbf{\emph{P}0} is nonconvex and thus difficult to straightforwardly minimize. To resolve this issue, we consider approximating the problem by dealing with the cases with small and large AoI exponents, respectively. Specifically, when the AoI exponent $\theta$ is small, the statistical AoI specified in Eq. (\ref{eq:effective-aoi_fading}) can be approximated by
\begin{align}\label{eq:effective-aoi_fading_small_theta}
\Delta_{\rm sta}^{s}(\theta) \approx 2T\left(\displaystyle\sum_{k = 1}^{K}\frac{p_k}{n_k}\right)^{-1}.
\end{align}
When $\theta$ is large, the statistical AoI can be approximated by
\begin{align}
\label{eq:effective-aoi_fading_large_theta}
\Delta_{\rm sta}^{l}(\theta) \approx \frac{1}{\theta}\log \sum_{k = 1}^{K}p_ke^{\displaystyle2\theta n_kT}.
\end{align}
Also, we have
\begin{align}\label{eq:error}
\left|\Delta_{\rm sta}^{l}(\theta) - \Delta_{\rm sta}(\theta)\right| \leq \frac{1}{\theta}\log n_{\rm max},\mbox{~for~}\forall \theta > 0.
\end{align}
Derivations of (\ref{eq:effective-aoi_fading_small_theta}) and (\ref{eq:effective-aoi_fading_large_theta}) are detailed as follows.
\subsubsection{The case with small $\theta$}
Since $T$ is generally small around the millisecond level (e.g., 1~ms in LTE-A and 5G), with a small $\theta$ we can have $e^{2\theta n_kT} \approx 1 + 2\theta n_kT$. Further, by using $\log(1 + x) \approx x$ for small $x$, Eq. (\ref{eq:effective-aoi_fading}) can be rewritten as
\begin{align}
\Delta_{\rm sta}(\theta)
&\approx \frac{1}{\theta}\log \sum_{k = 1}^{K}\mathrm{Pr}(A = 2n_kT)(1 + 2\theta n_kT) \\
&\approx \frac{1}{\theta}\log\left[1 + \sum_{k = 1}^{K}2\mathrm{Pr}(A = 2n_kT)\theta n_kT\right] \\
&\approx \sum_{k = 1}^{K}2\mathrm{Pr}(A = 2n_kT)n_kT. \label{eq:lemma_0_proof1}
\end{align}
Substituting Eq. (\ref{eq:pdf_aoi}) into (\ref{eq:lemma_0_proof1}), we obtain Eq. (\ref{eq:effective-aoi_fading_small_theta}).

\subsubsection{The case with large $\theta$}
Because $1\leq n_k\leq n_{\rm max}$, the PDF of peak age specified in Eq. (\ref{eq:pdf_aoi}) is bounded by
\begin{align}\label{eq:pdf_aoi_bounded}
 \frac{p_k}{n_{\rm max}}\leq \mathrm{Pr}(A = 2n_kT)\leq p_k n_{\rm max}.
\end{align}
Substituting Eq. (\ref{eq:pdf_aoi_bounded}) into (\ref{eq:effective-aoi_fading}), we have Eq. (\ref{eq:error}). When $\theta$ is large, the value of $\frac{1}{\theta}\log n_{\rm max}$ is very small. Thus, the approximation form specified in Eq. (\ref{eq:effective-aoi_fading_large_theta}) is derived. We should notice that $\Delta_{\rm sta}^{l}(\theta)$ does not correspond to the maximum of peak AoI. Only when $\theta = \infty$, we have $\Delta_{\rm sta}^{l}(\theta) = \max\{2n_1T,2n_2T, ..., 2n_KT\}$. When $\theta$ is finite, $\Delta_{\rm sta}^{l}(\theta)$ is capable of covering some cases of intermediate values of the AoI exponent.


\subsection{The Case of Small AoI Exponent}
The approximations of $\Delta_{\rm sta}(\theta)$, i.e., $\Delta_{\rm sta}^{s}(\theta)$ and $\Delta_{\rm sta}^{l}(\theta)$, decrease as $n_k$ decreases. Once the ``=" in Eq. (\ref{eq:bit_constraint}) does not hold, $n_k$ can be further reduced, which leads to the smaller $\Delta_{\rm sta}^{s}(\theta)$ and $\Delta_{\rm sta}^{l}(\theta)$. Thus, the optimal solution inevitably keeps ``=" in Eq. (\ref{eq:bit_constraint}) hold. Then, plugging Eq.~(\ref{eq:channel_capacity}) into Eq.~(\ref{eq:bit_constraint}), we can rewrite Eq.~(\ref{eq:bit_constraint}) as
\begin{align}\label{eq:transformer}
\mu_k = \frac{1}{\gamma_k}\left(\displaystyle 2^{g(n_k)} - 1\right),
\end{align}
where the function $g(n_k)$ is defined by
\begin{align}
g(n_k) \triangleq \frac{N_{\rm bit} + \sqrt{V n_kTB}f^{-1}_Q(\varepsilon)\log_2 e}{n_kTB}.
\end{align}
By substituting Eq. (\ref{eq:transformer}) into (\ref{eq:average_power_constraint}) and using the approximation form given in Eq. (\ref{eq:effective-aoi_fading_small_theta}), \textbf{\emph{P}0} can be converted to \textbf{\emph{P}1} as follows:
\begin{subequations}
\begin{align}
\textbf{\emph{P}1}:
\displaystyle \max_{n_k} \,\,\,& \sum_{k = 1}^{K}\frac{p_k}{n_k},\\
{\rm{s.t.}}\quad
& \sum_{k = 1}^{K}\frac{p_k}{\gamma_k}\left(2^{g(n_k)} - 1\right) \leq \bar{\mu},
\label{eq:sub_constraint_1}\\
& n_k \leq n_{\rm max}, \forall k \in \mathbb{K}. \label{eq:sub_constraint_2}
\end{align}
\end{subequations}
The objective of \textbf{\emph{P}1} is not clearly concave and thus it is difficult to derive the globally optimal solution via Lagrange method. Alternatively, \textbf{\emph{P}1} can be regarded as a concave-convex fractional programming problem. Then, the quadratic transform proposed in \cite{Shen2018Fractional}, which is a powerful way to cope with fractional optimization problems, can be applied. Specifically, based on \cite[Corollary 1]{Shen2018Fractional}, \textbf{\emph{P}1} is equivalently converted to $\textbf{\emph{P}1}^{(e)}$, which is given by
\begin{subequations}
\begin{align}
\textbf{\emph{P}1}^{(e)}:
\displaystyle \max_{n_k,m_k} \,\,\,& \sum_{k = 1}^{K}(2m_k\sqrt{p_k} - m_k^2n_k),\\
{\rm{s.t.}}\quad
& \mathrm{(\ref{eq:sub_constraint_1})} \mbox{~and~} \mathrm{(\ref{eq:sub_constraint_2})},
\end{align}
\end{subequations}
where $m_k$ is the introduced auxiliary variable. Then, we can solve $\textbf{\emph{P}1}^{(e)}$ by updating the following iterator until convergence:
\begin{subequations}
\begin{numcases}{\hspace{-1cm}}
m^{(i+1)}_k = \frac{\sqrt{p_k}}{n^{(i)}_k},\\
n^{(i+1)}_k = \arg \textbf{\emph{P}1}^{(e)}_{\rm sub}.
\end{numcases}
\end{subequations}
Here, $\textbf{\emph{P}1}^{(e)}_{\rm sub}$ is given as follows:
\begin{subequations}
\begin{align}
\textbf{\emph{P}1}^{(e)}_{\rm sub}:
\displaystyle \min_{n_k} \,\,\,& \sum_{k = 1}^{K}\left(m^{(i+1)}_k\right)^2n_k,\\
{\rm{s.t.}}\quad
& \mathrm{(\ref{eq:sub_constraint_1})} \mbox{~and~} \mathrm{(\ref{eq:sub_constraint_2})},
\end{align}
\end{subequations}
which is convex and can be solved via standard procedures and well-developed tools package.

\subsection{The Case of Large AoI Exponent}
When AoI exponent is large, by using the approximation form given in Eq. (\ref{eq:effective-aoi_fading_large_theta}), \textbf{\emph{P}0} can be transformed to \textbf{\emph{P}2} as follows:
\begin{subequations}
\begin{align}
\textbf{\emph{P}2}:
\displaystyle \min_{n_k} \,\,\,& \sum_{k = 1}^{K}p_ke^{2\theta n_kT},\\
{\rm{s.t.}}\quad
& \mathrm{(\ref{eq:sub_constraint_1})} \mbox{~and~} \mathrm{(\ref{eq:sub_constraint_2})}.
\end{align}
\end{subequations}
It can be readily verified that \textbf{\emph{P}2} is a convex problem. In the following, Lagrangian method is used. The Lagrangian function, denoted by $\mathcal{L}(n_k;\alpha,\beta_k)$, of \textbf{\emph{P}2} can be constructed as follows:
\begin{align}
\mathcal{L}(n_k;\alpha,\beta_k) = \sum_{k = 1}^{K}p_ke^{2\theta n_kT}
&\!+\! \alpha \left(\sum_{k = 1}^{K}\frac{p_k}{\gamma_k}\left( 2^{g(n_k)} \!-\! 1\right) \!-\! \bar{\mu}\right) \nonumber \\
&\!\!\!+\! \sum_{k = 1}^{K} \beta_k (p_k n_k \!-\! p_k n_{\rm max}),
\end{align}
where $\alpha$ and $\beta_k$ are the multipliers associated with Eqs. (\ref{eq:sub_constraint_1}) and (\ref{eq:sub_constraint_2}), respectively. Taking the first derivative of $\mathcal{L}(n_k;\alpha,\beta_k)$ with respect to $n_k$, we have
\vspace{-5pt}
\begin{align}
\frac{\partial\mathcal{L}(n_k;\alpha,\beta_k)}{\partial n_k} \!\!=\!\! 2p_k\theta T e^{2\theta n_kT} \!\!+\!\! \alpha \frac{p_k}{\gamma_k}g'(n_k)\displaystyle 2^{g(n_k)} \log 2 + \beta_k p_k, \nonumber
\end{align}
where $g'(n_k)$ is the first derivative function of $g(n_k)$. Then, the Karush-Kuhn-Tucker (KKT) conditions shall hold as follows:
\begin{subequations}\label{eq:kkt}
\begin{numcases}{\hspace{-1cm}}
2\theta T e^{2\theta n_kT} + \alpha \frac{g'(n_k)}{\gamma_k}2^{g(n_k)}\log 2+ \beta_k = 0, \label{eq:kkt_1}\\
\alpha\left(\sum_{k = 1}^{K}\frac{p_k}{\gamma_k}\left(2^{g(n_k)} - 1\right) - \bar{\mu}\right) = 0, \\
\beta_k ( n_k - n_{\rm max}) = 0, \forall k \in \mathbb{K}.\label{eq:kkt_3}
\end{numcases}
\end{subequations}
For description convenience, we denote by $n_k^*$ the solution to Eq. (\ref{eq:kkt}) and define the function $h(n_k, \gamma_k)$ as
\vspace{-6pt}
\begin{align}
h(n_k, \gamma_k) \triangleq 2\theta T e^{ 2\theta n_kT} \!\!+\! \alpha \frac{g'(n_k)}{\gamma_k}2^{ g(n_k)} \log 2.
\end{align}
The property of $n_k^*$ is described in Lemma \ref{lem:lemma_1}, which can facilitate to find out $n_k^*$.
\begin{figure*}
\vspace{-10pt}
\centering
\subfigure[]{\includegraphics[height=1.8in]{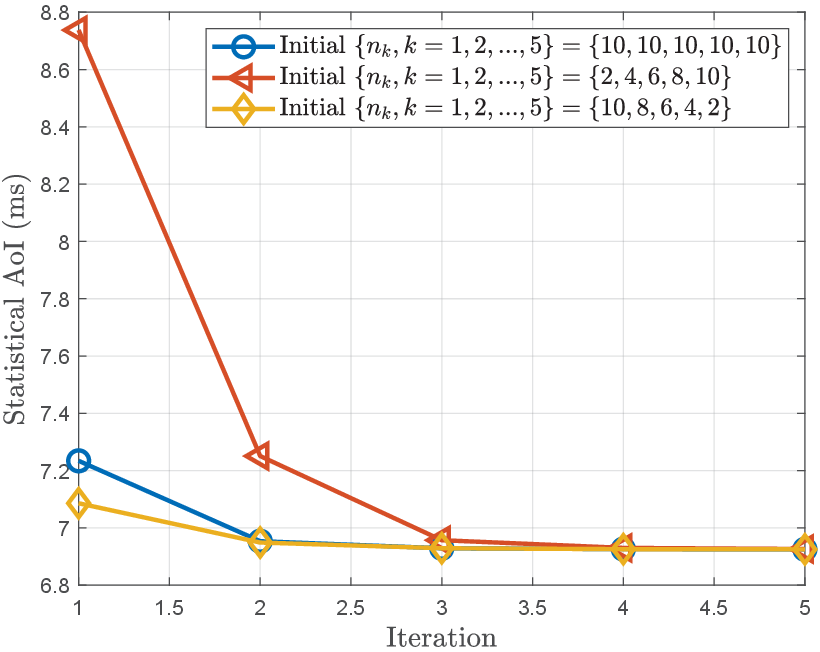}\label{fig:converage_performance}}
\subfigure[]{\includegraphics[height=1.8in]{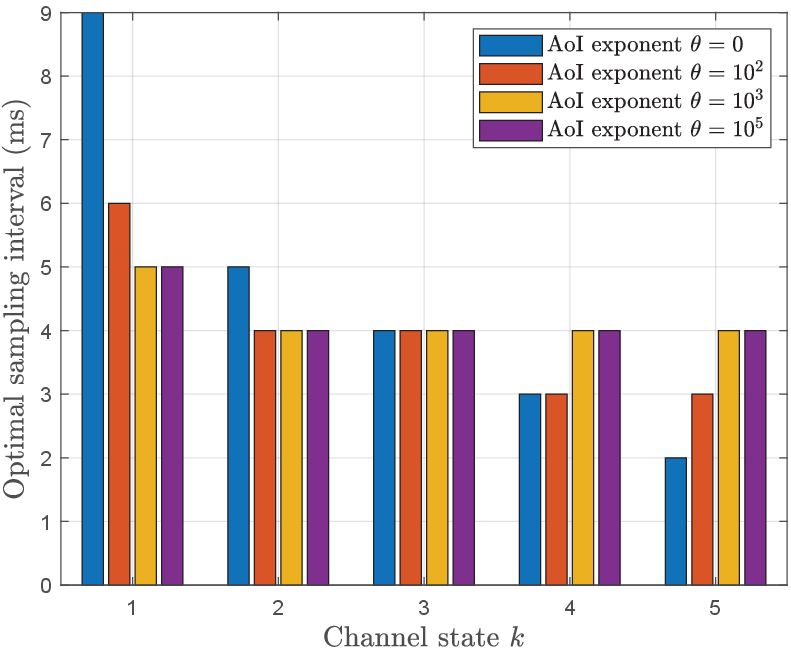}\label{fig:simulation_1}}
\subfigure[]{\includegraphics[height=1.8in]{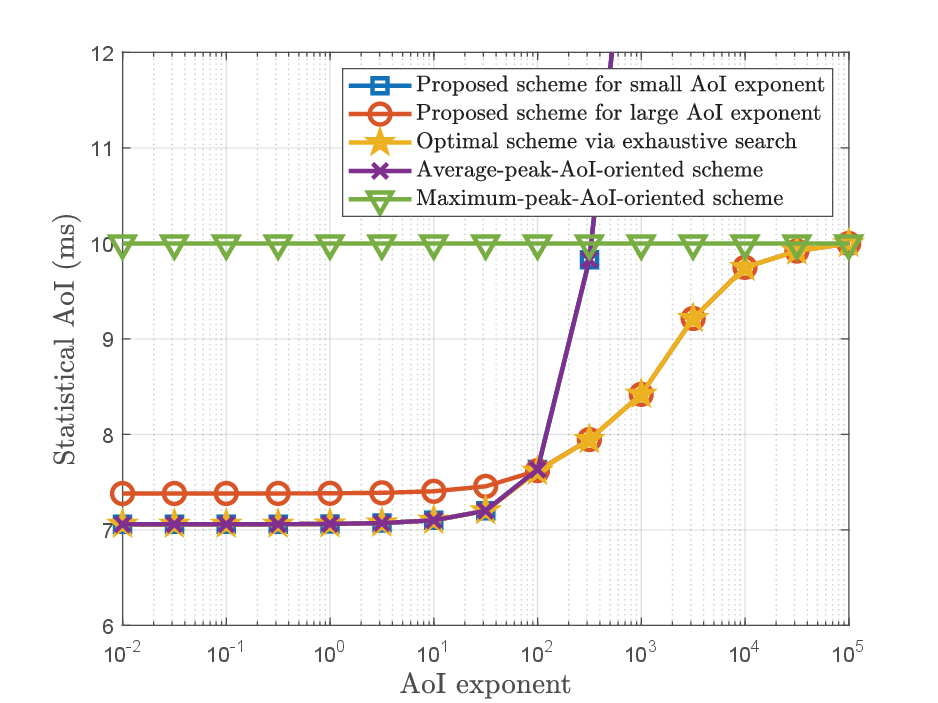}\label{fig:performance_eAoI}}
\label{fig:simulation_2}
\caption{(a) The convergence rate of using quadratic transform method to solve \textbf{\emph{P}1}, (b) our proposed scheme under different AoI exponents, and (c) the performance of our proposed scheme (Simulation value of each parameter: $N_{\rm bit}=500$ bits, $T=1$ ms, $B=50$ kHz, $n_{\rm max}=10$, $\gamma_k\in\{1,5,10,15,20\}$, $p_k\in\{0.1,0.2,0.4,0.2,0.1\}$, $P=1$ W, and $\varepsilon=10^{-3}$).\vspace{-20pt}}
\end{figure*}

\begin{lemma}\label{lem:lemma_1}
There exists $n_1^* \geq n_2^* \geq ... \geq n_K^*$ for arbitrary $\theta$.
In particular, when $\theta$ is finite, $n_k^*$ strictly decreases as $k$ increases unless $n_k^* = n_{\rm max}$. When $\theta$ is infinite, $n_k^*$ becomes a constant for all of $k$ in $\mathbb{K}$.
\end{lemma}
\emph{Proof}: Since $g'(n_k) < 0$,
the function $h(n_k, \gamma_k)$ is an increasing function of $\gamma_k$. Without loss of generality, we assume that $n_k^*$ is the first one among $\{n_{1}^*,n_{2}^*, ..., n_{K}^*\}$ strictly less than $n_{\rm max}$. In light of the slackness condition (\ref{eq:kkt_3}), we have $\beta_k = 0$. Then, based on Eq. (\ref{eq:kkt_1}), $h(n_k^*,\gamma_k) = 0$ can be obtained. Since $\gamma_{k+1} > \gamma_{k}$, we have $h(n_k^*,\gamma_{k+1}) > 0$. Also, $h(n_k, \gamma_k)$ strictly increases as $n_k$ increases for arbitrary finite $\theta$. To guarantee $h(n_{k+1}^*, \gamma_{k+1})$ equal to zero, we should have $n_{k+1}^* < n_{k}^*$. Till now, we have validated that for arbitrary finite $\theta$, $n_1^* \geq ... \geq n_k^* > n_{k + 1}^* > ... > n_K^*$. When $\theta\rightarrow\infty$, $h(n_k, \gamma_k)$ increases as $n_k$ increases. Thus, we still have $n_1^* \geq n_2^* \geq ... \geq n_K^*$, and assume that $n_k^*$ is the first one strictly less than $n_{\rm max}$, resulting in $\beta_k = 0$. Then, Eq. (\ref{eq:kkt_1}) can be rewritten as follows:
\begin{align}
\frac{2\theta T}{\alpha} e^{2\theta n_kT} = - \frac{g'(n_k)}{\gamma_k}2^{g(n_k)}\log 2.
\end{align}
Performing $(\cdot)^{\frac{1}{2\theta T}}$ and letting $\theta\rightarrow\infty$, we have
\begin{align}
\left(\frac{2\theta T}{\alpha}\right)^{\frac{1}{2\theta T}} e^{n_k} = 1 \hspace{.2cm}\Rightarrow\hspace{.2cm} n_k^* = \lim_{\theta\rightarrow\infty} \log \left(\frac{\alpha}{2\theta T}\right)^{\frac{1}{2\theta T}},
\end{align}
implying that $n_k^*$ is independent of $\gamma_k$. Note that the value of $\alpha$ will be also close to the infinite when $\theta$ is infinite.
$\hfill\blacksquare$

Based on Lemma \ref{lem:lemma_1}, we introduce a variable $k^*$ and suppose that $n_k = n_{\rm max}$ for $\forall k < k^*$, and $n_k < n_{\rm max}$ for $\forall k \geq k^*$. During each status update, since the size of one status packet is fixed, the more slot scheduled, the less power consumed. As $k^*$ increases, more $n_k$ will be assigned approaching to $n_{\rm max}$, which assures small average power consumption. Thus, we can use bisection search to find the optimal $k^*$. Further, since function $h(n_k, \gamma_k)$ is decreasing as $\alpha$ increases, we can use bisection search to find the optimal $\alpha$. In addition, because $h(n_k, \gamma_k)$ is increasing with larger $n_k$, the bisection search can be effectively employed to identify the optimal $n_k$. More details are omitted here for lack of space. Since one-dimensional bisection search is often used, the low computational complexity can be guaranteed.

The above-derived $n_k$ is generally a decimal, however, the variable $n_k$ should be an integer. To convert it to an integer, we substitute the optimal $n_k$ with either its rounded up or rounded down equivalent. For example, when the optimal $n_k$ is 6.3, we replace it with either 7 or 6. Finally, we find the optimal solution by rounding each $n_k$ either upwards or downwards, which should be validated to meet the constraints of the problem. The corresponding computational complexity is $O(2^K)$ far less than that of the exhaustive search method, i.e., $O(n_{\rm max}^K)$. For a specified AoI exponent, we select the scheme that can achieve the smaller statistical AoI between the schemes for the small and large AoI exponents as the final sampling internal scheme.



\section{Numerical Results}
In this section, we evaluate the performance of our proposed sampling interval scheme for status update transmission over multi-state fading channels.
Figure \ref{fig:converage_performance} shows the performance of using the quadratic transforming method to solve the approximated statistical AoI minimization problem for the case of small AoI exponent. We can observe that under the different initial $\{n_k, k = 1,2,...,K\}$, the minimum statistical AoI can always be reached about 5-round iterations.


Figure \ref{fig:simulation_1} shows the optimal sampling interval scheme under different AoI exponents. It can be seen from the figure that when AoI exponent is small, the optimal sampling interval is decreasing as the CSI increases, and when the AoI exponent is infinite, the optimal sampling interval is nearly independent of the CSI. This result is aligned with Lemma~\ref{lem:lemma_1}.

Figure \ref{fig:performance_eAoI} shows the performance of our proposed scheme under different AoI exponents. The performance of optimal scheme derived by using the time-inefficient exhaustive search method is also presented. It can be seen that the performance of our proposed scheme is quite close to the optimal scheme. Thus, our proposal can achieve nearly the highest level of information freshness under the various AoI exponents. Moreover, the performance of the average-peak-AoI-oriented scheme and the maximum-peak-AoI-oriented scheme is assessed, respectively. We can see that only when the AoI exponent is small or infinite, the average-peak-AoI-oriented scheme or the maximum-peak-AoI-oriented scheme can achieve the minimum statistical AoI.
\section{Conclusions}
We investigated the statistical AoI minimization problem for status update over multi-state fading channels, which was approximated to a fractional programming problem and a convex problem for small and large AoI exponents, respectively. Solving them, We derived an efficient and near-optimal sampling interval scheme, in which how the sampling interval should be tuned with different AoI exponents was shown. In particular, for the extremely stringent AoI requirement, the sampling interval converged to a constant regardless of CSI's variation. The numerical results verified that our proposed scheme can significantly reduce the AoI with the given age violation probability and thus guarantee the excellent performance in terms of freshness of information. In future work, we would like to extend the research to multi-status statistical-AoI optimization in various multiple access scenarios, such as TDMA, FDMA, and NOMA. Moveover, our work can be seamlessly incorporated with some wireless transmission techniques, such as intelligent reflected surface techniques, to enhance the channel quality for providing higher level of information freshness.

\bibliographystyle{IEEEtran}
\bibliography{References}

\end{document}